\documentclass[12pt,preprint]{aastex}
\usepackage{amssymb}
\usepackage{graphicx}
\usepackage{epsfig}
\usepackage{graphicx}
\usepackage{latexsym}

\begin{document}

\title{A new class of gamma-ray bursts from stellar disruptions by intermediate mass black holes}

\author{H. Gao\altaffilmark{1,2}, Y. Lu\altaffilmark{1}, S. N. Zhang\altaffilmark{1,3}}
\altaffiltext{1}{National Astronomical Observatories, Chinese
Academy of Sciences, Beijing 100012, China, gaohe@mail.bnu.edu.cn}
\altaffiltext{2}{Department of Astronomy, Beijing Normal University,
Beijing 100875, China}
\altaffiltext{3}{Key Laboratory of Particle Astrophysics, Institute
of High Energy Physics, Chinese Academy of Sciences, Beijing 100049,
China}

\begin{abstract}
It has been argued that the long gamma-ray burst (GRB) of
GRB\,060614 without associated supernova (SN) has challenged the
current classification and fuel model for long GRBs, and thus a
tidal disruption model has been proposed to account for such an
event. Since it is difficult to detect SNe for long GRBs at high
redshift, the absence of an SN association cannot be regarded as the
solid criterion for a new classification of long GRBs similar to
GRB\,060614, called GRB\,060614-type bursts. Fortunately, we now
know that there is an obvious periodic substructure observed in the
prompt light curve of GRB\,060614. We thus use such periodic
substructure as a potential criterion to categorize some long GRBs
into new class bursts, which might have been fueled by an
intermediate-mass black hole (IMBH) gulping a star, rather than a
massive star collapsing to form a black hole. Therefore, the second
criterion to recognize these new class bursts is if they fit the
tidal disruption model. From a total of 328 \emph{Swift} GRBs with
accurate measured durations and without SN association, we find 25
GRBs satisfying the criteria for GRB\,060614-type bursts: 7 of them
are with known redshifts, and 18 with unknown redshifts. These new
bursts are $\sim 6\% $ of the total \emph{Swift} GRBs, which are
clustered into two subclasses: Type I and Type II with considerably
different viscous parameters of accretion disks formed by tidally
disrupting their different progenitor stars. We suggest that the two
different kinds of progenitors are solar-type stars and white
dwarfs: the progenitors for 4 Type I bursts with viscous parameter
of around 0.1 are solar-type stars, and the progenitors for 21 Type
II bursts with viscous parameter of around 0.3 are white dwarfs.
Potential applications of this new class of GRBs as cosmic standard
candles are discussed briefly.

\end{abstract}


\section{Introduction}
Recently the peculiar long-duration GRB\,060614 poses a great
challenge to the widely accepted concept that long gamma-ray bursts
(GRBs) are the consequences of core collapse of very massive stars.
GRBs are normally classified into two groups: the long-duration
bursts ($T_{90} > $ 2 s) and short ones ($T_{90} <$ 2 s)
\citep{Kou93}, where $T_{90} $ is defined as the time interval in
which the integrated photon counts increase from 5\% to 95\% of the
total counts. Several nearby long GRBs were observed to be firmly
associated with core-collapse supernovae (SNe), and meanwhile the
``collapsar'' model of massive star explosions leading to long GRBs
has been well developed (\citealt{Woo06}). On the other hand, short
GRBs are hypothesized to be formed by the coalescence of binary
compact stars, and hence with no SN connection (\citealt{Nak07}).
The prompt light curve of GRB\,060614 as detected by the
\emph{Swift} Burst Alert Telescope (BAT) instrument displayed an
initial short spike lasting for $\sim 4$\,s followed by an extended
component lasting for $\sim 100$\,s, the latter being softer than
the former in energy. This is a long burst by the above definition.
The measured redshift of its host galaxy is as low as 0.125
\citep{Pri06}. However, despite its promising proximity,
surprisingly no SN was observed to accompany the GRB down to very
deep detection limits \citep{Del06a,Fyn06,Gal06}.

Various solutions have been proposed for the missing-SN puzzle of
GRB\,060614: First, the GRB perhaps did not take place at
$z\!=\!0.125$ at all, but at much higher redshift so that its
associated SN was below the detection limit. \citet{Cob06} claimed
that the proximity of the GRB line of sight to the $z\!=\!0.125$
galaxy was a chance coincidence; however \citet{Cam08} found that
the chance coincidence probability is less than 0.02\%, thus ruling
out the chance coincidence assumption with high confidence.
Alternatively, the GRB could be produced with a very faint
core-collapse SN \citep{Tom07,Fry07}, similar to the ones described
by \citet{Tur98}, \citet{Pas04,Pas07} and \citet{Val09}. Finally,
this long GRB was not a consequence of the core-collapse of a
massive star, and hence a novel mechanism is needed. For example,
\citet{Kin07a} suggested that the merger of a massive white dwarf
with a neutron star can make a long-duration GRB.

Recently, \cite{Lu08} proposed a new mechanism to produce GRBs,
namely the tidal disruption of a star by an intermediate mass black
hole (IMBH). In this scenario, both the long duration and the lack
of an associated SN, as observed in GRB\,060614, are naturally
expected. Furthermore, the model can well explain a probable 9-s
periodicity found in the prompt BAT light curve of GRB\,060614
between 7 and 50 s. Such a substructure seems difficult to fit into
either the collapsar scenario or compact star mergers. It is thus
natural to investigate if GRB\,060614 is just the first of this new
class of GRBs, which is the purpose of this paper.

We focus on the periodic substructure in the prompt light curve of a
burst, since for most GRBs it is impossible to tell if an SN is
associated or not, due to their high redshifts. A general picture of
the tidal disruption model is given in Section 2, where important
physical quantities are defined. The selection criteria, and the
selected GRB\,060614-like events in the \emph{Swift} samples, are
described in Section 3. We do statistical studies on these bursts,
and combine the statistical results with the model to give
predictions in Section 4. Our conclusion and discussions are given
in Section 5.

\section{The Tidal Disruption Model for gamma-ray bursts}

\subsection{The model description}

Tidal disruption of a star by an IMBH is analogous to the case for a
supermassive black hole. A star, which happened to be close enough
to the black hole, was distorted and squashed into a pancake by the
strong tidal forces of the black hole. Once the star is tidally
disrupted by the black hole, the squashed debris finally falls into
the black hole's horizon, forming a transient accretion disk around
the black hole. During the early high accretion rate stage (near the
Eddington rate), the inner region of the disk should be dominated by
radiation pressure. In this case, the disk within the spherization
radius $R_{\rm sp}$ is thermally unstable \citep{Sha73}, and the
material in this inner region is likely broken into many blobs. When
the blobs are dragged into the black hole, the seed magnetic field
anchored in the blobs can be amplified, forming a strong and ordered
poloidal field, which in turn threads the black hole with a
mass-flowing ring in the inner region of the disk and extracts a
large amount of rotational energy, creating two counter-moving jets
along the rotation axis of the black hole \citep{Bla77}. As in the
conventional GRB model, each jet pointed toward the observer,
produces one mini-burst lasting over a blob's free-falling timescale
\citep{Lu08}. Consequently, many mini-bursts should be produced for
the tidal disruption event. Assuming that the in-falling process of
the blobs into the black hole is neither uniform nor completely
unsystematic, they may fall in groups quasi-periodically and this
behavior should be modulated by the Keplerian timescale, forming a
periodic sub-burst. All of these mini-bursts in the sub-burst add
together to form a GRB, and the duration of the GRB is determined by
the time when all the blobs with the spherization radius are removed
and fall into the black hole at the marginally stable radius. Note
that the blobs in each group randomly fall into the black hole, so
small dispersion between sub-bursts' durations may exist.This model
can reasonably explain all the observed basic features of the
unusual GRB\,060614, including the duration, the total energy, the
periodic substructure, and most importantly, the absence of SN link
\citep{Lu08}. The general picture of the tidal disruption model for
such a GRB is plotted in Figure 1.

For convenience, we introduce the following dimensionless quantities
throughout this paper:
$$M_5=\frac{M_{\rm bh}}{10^5M_\odot}, \dot{m}=\frac{\dot{M}}{\eta
\dot{M}_{\rm Edd}},$$ where $\dot{M}_{\rm Edd}=3\times
10^{-2}\eta_{0.1}^{-1}M_5\,M_\odot\,{\rm yr}^{-1}$ is the Eddington
accretion rate, $\eta$ is the energy conversion factor and
$\eta_{0.1}=\eta/0.1$. Subsequently, the key physical parameters
related to the tidal disruption model derived by \citet{Lu08} are
briefly described.

\subsection {The three timescales and one relation}

According to \citet{Lu08}, there are three useful timescales related
to explain the observations of GRB\,060614: the mini-burst duration,
$T_{\rm pulse}$, the sub-burst period, $T_k$, and the duration of
the whole GRB, \textbf{$T_{90}$}. For instance, the three timescales
are marked in the prompt light curve of GRB\,060614 in Figure 2.
They can be calculated by
\begin{eqnarray}
&& T_{pulse}\simeq 3M_5\, {\rm s} \,, \\
&& T_{k}\simeq50 \hat{r}_{\rm ms}^{3/2}M_5 \, {\rm s} , \\
&& T_{90}\simeq 50\alpha^{-1}\hat{r}_{\rm ms}^{3/2}M_5\, {\rm s} \,,
\end{eqnarray}
where $\alpha$ is the viscous parameter,  $0\leq\alpha\leq 1$, and
$\hat{r}_{\rm ms}$ is the dimensionless radius of the marginally
stable circular orbit around a non-spinning in units of $6GM/c^2$,
i.e., $\hat{r}_{\rm ms}=1$.

From Equations (2) and (3), we can obtain a linear relation between
the duration of the bursts and the period of the substructure
\begin{eqnarray}
T_k=\alpha T_{90}\,\,.
\end{eqnarray}
Equation (4) shows that the slope of the linear relation corresponds
to the viscous parameter of the disk. This relation could be very
useful for the classification of the GRBs.

Nevertheless, the viscous parameter, $\alpha$, is uncertain, which
is considered to be related to the structure and physical properties
of the disk, especially the magnetic fields of the disk; recent
numerical studies of the magnetorotational instability viscosity
mechanism (Balbus \& Hawley 1991) have shown that the value of the
viscosity parameter depends upon the magnetism of the disk and
sufficiently strong magnetic fields in the disk are necessary for a
large viscosity parameter (e.g., $\alpha>1$; Pessah et al. 2007).
Note that the disk considered here is formed through the black hole
gulping a star, indicating that the disruption of a different star
by black holes will give a different disk, then the different
viscosity of the disk. To apply Equation (4) to certain GRBs,
different values of $\alpha$ indicate different GRB subclasses,
depending on the progenitors of the bursts.

\subsection {The energy for a GRB}

The tidal disruption model \citep{Lu08} can predict the isotropic
energy of a GRB, $E_{\rm iso}$, given the beaming factor of
$\Gamma$,
\begin{eqnarray}
E_{\rm iso} \simeq E_{\rm tot} \times \Gamma,
\end{eqnarray}
where $10\leq\Gamma\leq 1000$, and $E_{\rm tot}$ is the total energy
of a GRB, which is from the rotational energy of black holes
extracted by the BZ process \citep{Bla77},
\begin{eqnarray}
E_{\rm tot}\simeq 2.46\times
10^{51}\alpha^{-1}M_5\left[\frac{(2.52\dot{m})^{7/64}-1}{2.52\dot{m}-1}\right]A^2f(A)N_{\rm
tot}\,\, {\rm erg},
\end{eqnarray}
where $N_{\rm tot}$ is the total number of mini-bursts in the whole
GRB, $A$ is the dimensionless angular momentum of the black hole,
and $f(A)=2/3$ for $A\rightarrow 0$ and $f(A)=\pi-2$ for
$A\rightarrow 1$ \citep{Lee00}).

Substituting Equation (4) into Equation (6) to eliminate the viscous
parameter $\alpha$, the expression of $E_{\rm iso}$ can be rewritten
as
\begin{eqnarray}
E_{\rm iso}\simeq 2.46\times
10^{51}\frac{T_{90}}{T_k}\left[\frac{(2.52\dot{m})^{7/64}-1}{2.52\dot{m}-1}\right]\Gamma
M_5A^2f(A)N_{\rm tot}\,\, {\rm erg}\,.
\end{eqnarray}
Given $A^2f(A)=0.02$ and $\dot{m}=1$, \citet{Lu08} consistently
explained the properties of GRB\,060614. Since the purpose of this
work is searching for GRBs similar to GRB\,060614 and then analyzing
these bursts' properties, it is reasonable to assume that the
adoption of $A^2f(A)=0.02$ and $\dot{m}=1$ is still suitable.
Therefore, we use these values to other bursts hereafter in this
paper. Note that $E_{\rm iso}$ is the most important physical
quantity for using GRBs to investigate cosmology, such as the
relation of \citet{Ama02}. Equation(7) shows that the isotropic
energy only depends on the prompt gamma-ray emission light curve of
a burst and the beaming factor of each individual burst, if the
tidal disruption model for GRB\,060614 can be generalized to other
bursts similar to GRB\,060614. This is in favor of $E_{\rm iso}$ of
each burst as a calibrated candle to study cosmology. We will
discuss this later.

\section{Searching for gamma-ray bursts similar to GRB060614}

\subsection{The criteria and Samples}
Assuming that the tidal disruption model for GRB\,060614 can be
generalized to investigate other gamma-ray bursts similar to
GRB\,060614, called GRB\,060614-type bursts, we propose three
criteria to identify this new class of bursts: (1) the bursts must
be long-duration bursts ($T_{90}>2$~s) without SN association; (2)
the bursts should have an obvious periodic substructure (sub-burst)
in the prompt light curves of the bursts, and the periodic
substructure is composed of more than three mini-bursts; and (3) the
burst should satisfy the relation indicated by the tidal disruption
model: $T_{k}=\alpha T_{90}$. The three criteria must be all
satisfied for each burst identified.

According to the above criteria, we select the samples among the 393
GRBs discovered by the \emph{Swift} BAT before 2008 October 1. There
are 330 long-duration bursts with accurately measured $T_{\rm 90}$.
For these 393 GRBs, we first exclude two GRBs with established SN
associations: GRB060218 \citep{Pia06} and GRB050525A \citep{Del06b}.
From the rest 328 GRBs, we identified 24 new bursts. The prompt
light curve of GRB\,060614 is shown in Figure 2 and the light curves
of the newly selected 24 samples are shown in Figure 3. All of them
are from a public online database
\footnote{http://grb.physics.unlv.edu/~xrt/xrtweb/web/sum.html}.
Tables 1 and 2 list all the physical quantities related to the 25
samples (including GRB\,060614), where Table 1 corresponds to 7
samples with measured redshifts, and Table 2 is for the rest 18
samples with unknown redshifts. The data are derived by combining
the observations and the model. The details are summarized as
follows.

\subsection{The observed quantities}

(1) From a public online database
\footnote{http://swift.gsfc.nasa.gov}, we can obtain the duration
($T_{90}$) of the 25 GRB\,060614-type bursts, and the redshifts
($z^{\rm obs}$) of 7 GRB\,060614-type bursts including GRB 060614
itself.

(2) Based on the appearance of the observed light curves of the
bursts shown in Figures 2 and 3, we extract the number, $N_{\rm
sub}$, and the duration, $T_{k}$, of the sub-bursts. First our
attempt of power density spectrum (PDS) analysis failed in this
process. Taking GRB\,070223 as an example, whose light curve (see
Figure 3) shows obvious quasi-periodic substructures, we present its
PDS (the analysis is based on a 64 ms BAT light curve data, 15-350
keV and $T_{\rm -20}$\,s to $T_{\rm +110}$\,s ) in Figure 4. No
significant signal was found in the PDS. We then propose two
explanations for this failure. First, the extreme complexity of the
light curves and the sensitive dependence of PDS on the noise
properties sometimes prevent the PDS analysis from detecting the
obvious structures in the light curves; second, as we said in
Section 2, the scatter of sub-bursts' durations may lower the
detection significance of the sub-burst structure in the PDS.
Consequently, we obtain the results of sub-bursts by a visual
inspection method. In the following, we describe the detailed
process of identifying sub-bursts and estimating their durations
from the light curves of the 25 GRB\,060614-type bursts by the
visual inspection method. We want to stress two important features
of the GRB\,060614-type bursts' light curves that will be taken as
our selection \emph{priors}: (1) pulses in the light curves tend to
gather into several groups which are considered as candidates for
sub-bursts; and (2) quasi-periodic substructures exist in the light
curves. The 25 light curves are consequently classified into three
cases (details are presented in Tables 1 and 2). Case I ($2/25$):
all of the candidates for sub-bursts are separated by quiescent
periods (since our analysis is based on existing light curves whose
backgrounds have been taken out, we define quiescent period here as
the period during which no significant peak can be recognized
visually). Case II ($19/25$): all of the candidates for sub-bursts
join together. Case III ($4/25$): some of the candidates for
sub-bursts are separated by quiescent periods and others join
together. For Case I, $N_{\rm sub}$ is easily obtained. We then take
the average value of sub-bursts' durations as $T_{k}$. For Case II,
we first find out the highest peak in each sub-burst candidate, and
then find out the minimum of the valley between the two adjacent
highest peaks. These valleys, together with the beginning of the
first sub-burst candidate and the end of the last sub-burst
candidate, are taken as the candidates for the beginning or end of
these candidate sub-bursts, whose corresponding times are labeled as
$t_i\,(i=1\ldots n+1)$, where $n$ is the number of candidate
sub-bursts. Since the real sub-bursts should satisfy the
quasi-periodic property requirement, the values of all $t_{i+1}-t_i$
should be approximately the same. We calculate $T=\langle
t_{i+1}-t_i\rangle$ as the first estimate to $T_{k}$. For each
sub-burst candidate, if $t_{i+1}-t_i<0.5T$ or $t_{i+1}-t_i>1.5T$, we
reject the sub-burst candidate corresponding to $t_{i+1}$, and
repeat the above procedure with the remaining $n-1$ candidates until
all the candidates satisfy the criterion. Finally, we obtain $N_{\rm
sub}$ and take the value of $T=\langle t_{i+1}-t_i\rangle$ as
$T_{\rm k}$. For Case III, some sub-bursts could be first identified
similar to Case I, and we take the average value of these
sub-bursts' durations $T$ as the first estimate to $T_{k}$, and then
identify other sub-bursts through the same procedure for Case II.
Finally we also take the average value of the durations of these
sub-bursts as $T_{k}$ in Case III. We note that in all the cases, we
cannot give the exact error of $T_{k}$ introduced by the visual
inspection method.

(3) Following the peak-finding algorithm proposed by \citet{Li96},
we obtain the total number of mini-bursts, $N_{\rm tot}$, and the
corresponding timescale of $T_{\rm pulse}$. We use a linear function
\emph{B(t)} to fit the burst background between the pre-burst and
post-burst regions, where $t$ is the time. The whole burst region is
divided into many count bins. Corresponding to the mini-burst time,
$t_{\rm pulse}$, there is a peak bin with a count of $C_p$. This
peak bin is assumed as having more counts than the neighboring bins
around it. The condition for $C_p$ is satisfied if
\begin{eqnarray}
C_p - C_{\rm 1,2} \geq N_{\rm var} \sqrt{C_p}\,\,,
\end{eqnarray} where
$N_{\rm var}$ is a constant parameter, and $N_{\rm var} = 5$, $C_1$
and $C_2$ are the counts in two of the neighboring bins at $t_1$ and
$t_2$, where $t_1<t_{p}<t_2$ (see \citep{Li96} for details of the
peak-finding algorithm). Searching through the whole burst light
curve, all peaks can be found. Assuming that each peak count
corresponds to a mini-burst, the total mini-burst number, $N_{\rm
tot}$, in each GRB can be determined immediately. Combining the
assumption of the tidal disruption model, we can get
\begin{eqnarray}
T_{pulse}=\frac{N_{sub}}{N_{tot}}T_k\,\,,
\end{eqnarray}
where $T_k$, $N_{\rm sub}$ and $N_{\rm tot}$ are derived as in the
above discussion.

(4) The isotropic energy for seven known redshift bursts, $E^{\rm
KB08}_{\rm iso}$, are from \citet{Koc08}. This will be used as
evidence to check the tidal disruption model by comparing with the
model predictions.

\subsection{The model quantities}
In this subsection, we address and generalize the physical
quantities predicted by the tidal disruption model for the
GRB\,060614
 \citep{Lu08}, such as, the masses of black holes, the isotropic energy, and the
redshifts related to all 25 samples.

(1) The existence of IMBHs is still hotly debated and for which
astronomers are still searching for direct evidence \citep{Heg02,
Por04, Mil02, Geb05}. Recent theoretical work, however, has given
their possible density and occurring rate in the Milky Way. If these
predictions are confirmed by observations, there could be
1,000-10,000 IMBHs in our native Galaxy. One way to estimate the
masses of IMBHs is the tidal disruption model for GRBs. From
Equation (2), we find that the periodic timescale, $T_{k}$, is
linearly related to the masses of black holes assuming $\hat{r}_{\rm
ms}=1$ \citep{Lu08}. Rewriting Equation (2), we have
\begin{eqnarray}
M_{\rm 5} \simeq 0.02 T_{k}.
\end{eqnarray}
Substituting $T_k$ estimated according to the burst observations (in
Subsection 3.2) into Equation (10), we can immediately obtain $M_5$
for the 25 samples, which are listed in Tables 1 and 2. The results
show that the masses of the black holes range from $5\times
10^3M_\odot$ to $9\times 10^4M_\odot$. These are the typical masses
of IMBHs discussed by \citet{Mil02}. The distribution of $M_5$ for
the 25 GRB\,060614-type bursts is plotted in the left panel of
Figure 5.

(2)Adopting $A^2f(A)=0.02$, $\dot{m}=1$(see Section 2 for the
definition of $A^2f(A)$ and $\dot{m}$), and substituting $M_5$,
$T_{90}$ and $T_k$ into Equation (7), we can immediately calculate
the isotropic energy, $E_{\rm iso}^{\rm pre}$ as long as we know the
beaming factors for each GRB. Since we have not obtained the beaming
factors for all the 25 samples, for rough estimation of $E_{\rm
iso}^{\rm pre}$, we uniformly take the average value of 500 for
$\Gamma$, whose observational range is $10\leq\Gamma\leq 1000$, as
the beaming factor for all the 25 samples. The results are listed in
Tables 1 and 2. The middle panel of Figure 5 plots the distribution
of $E_{\rm iso}^{\rm pre}$ for the 25 samples.

(3) The redshifts for the 25 samples can also be predicted. It is
known that the relation between the isotropic energy, $E_{\rm iso}$,
and the fluence, $F_{\rm \gamma}$, of GRBs, satisfies
\begin{eqnarray}
\label{Eiso} E_{\rm iso} = 4\pi d_{\rm L}^{\rm 2}F_{\rm \gamma},
\end{eqnarray}
where $d_{\rm L}$ is the distance between a GRB and the observer,
which can be calculated by \citep{Car92}
\begin{eqnarray}
\label{dl} d_{\rm L} = \frac{1+z}{H_{\rm 0}}\int_0^z{[(1+x)^{\rm
2}(1+x\Omega_{\rm M})-x(2+x)\Omega_{\rm \Lambda}]^{-1/2}}dx,
\end{eqnarray}
where $x$ is an integral variable for the redshift; $\Omega_{\rm
M}$, $\Omega_{\rm \Lambda}$ and $H_{\rm 0}$ are the cosmological
constant parameters and Hubble constant, respectively. To compare
with the $E_{\rm iso}^{\rm KB08}$, we adopt $\Lambda$CDM cosmology,
that is: $\Omega_{\rm M} = 0.3$ , $\Omega_{\rm \Lambda} = 0.7$ and
$H_{\rm 0} = 71 {\rm km s}^{\rm -1}{\rm Mpc}^{\rm -1}$. Combining
Equations (11) and (12), we can numerically calculate the redshift,
$z$. The distribution of the predicted redshifts for 25 samples is
plotted in the right panel of Figure 5.

\section{Statistical relations and predictions}

Based on observations and our theoretical model, Subsections 2.2 and
2.3 have considered the physical quantities related to the 25
GRB\,060614-type bursts. The statistical analysis and the
predictions based on these quantities will be addressed in detail.

It is important to examine if the tidal disruption model, which is
successful for the case of GRB 060614 \citep{Lu08}, can be
generalized to explain the other 24 bursts selected in this paper.
We first consider the relation between $E_{\rm iso}^{\rm BK08}$ and
$E_{\rm iso}^{\rm pre}$ for the seven known redshift samples. We
plot the data of $E_{\rm iso}^{\rm BK08}$ and $E_{\rm iso}^{\rm
pre}$ in Figure 6. Using the least-squares fitting, we find a linear
relation
\begin{eqnarray}
 \log(E_{\rm iso}^{\rm KB08})=(0.997\pm
 0.03)\log(E_{\rm iso}^{\rm pre})\,\,,\nonumber
\end{eqnarray}
where the adjusted $\Re$-square value is $\sim 0.999$. This
indicates that the isotropic energy predicted by the model agrees
well with those given by the observations \citep{Koc08}. We thus
believe that the tidal disruption model can be successfully
generalized to study the properties of the other GRB\,060614-type
bursts. More observed isotropic energy for the rest 18 unknown
redshift GRB\,060614-type bursts can be used to check such relation
in the future if their host galaxy's redshifts can be measured.

It is interesting to note that the isotropic energy of GRBs is one
of the best rulers to measure the expansion of the universe
\citep{Ama02}. Since $E_{\rm iso}^{\rm pre}$ predicted by the tidal
disruption model is only related to $T_{90}$ and $T_k$, we can thus
study cosmology using only the prompt light curves of the bursts and
the redshifts of their host galaxies.

We then plot the data of $z^{\rm obs}$ and $z^{\rm pre}$ for the
seven known redshift samples in Figure 7. The relation fitted by the
least-squares method follows:
\begin{eqnarray}
z^{\rm obs}=(1.10\pm 0.12)z^{\rm pre}\,,\nonumber
\end{eqnarray}
where the adjusted $ \Re$-square is $0.93$. This again indicates
that the redshifts predicted agree well with those observed, which
further prove that the tidal disruption model is correct for the new
class of bursts, such as GRB\,060614-type bursts discussed in this
paper. The predictions of the 18 unknown redshift GRB\,060614-type
bursts will be tested by the further observations.

The relations of both $E_{\rm iso}^{\rm pre}-E_{\rm iso}^{\rm KB08}$
and $z^{\rm pre}-z^{\rm obs}$ argue that the tidal disruption model
can work well for the 25 new class bursts. We plot the data of
$T_{k}$ and $T_{\rm 90}$ in Figure 8. We find interestingly that the
relation between $T_{k}$ and $T_{\rm 90}$ for the 25 samples can be
well fitted by two linear curves with different slopes: four of the
samples , i.e., GRB060210, GRB060614, GRB080602, and GRB080503,
follow the linear relation
\begin{eqnarray}
T_{k} = (0.08\pm 0.003)~T_{\rm 90}\,,
\end{eqnarray}
and the rest 21 of the samples follow
\begin{eqnarray}
T_{k} = (0.27\pm 0.013)~T_{\rm 90} \,,
\end{eqnarray}
respectively. The slope of the former curve is about $0.08$, and the
later one is $0.27$.

Comparing with Equations (13) and (14) with Equation (4), we get two
values of viscous parameters, $\alpha_1=0.08$ and $\alpha_2=0.27$,
respectively. Both of them fall well in the typical range of
$\alpha\sim$ 0.1-0.4 inferred from observations of unsteady
accretion disks, such as in the outbursts of dwarf novae and X-ray
transients \citep{Kin07b}. It is known that the viscous parameter
depends on the detailed structure and the magnetic field of the
disk, because more strongly magnetized allows more efficient angular
momentum transfer in the disk, and thus results in larger values of
$\alpha$ (Pessah et al. 2007). Here in our model the transient disk
is formed by the IMBHs disrupting stars, thus different structure
and magnetic field of the disk will be produced by the disruptions
of different type of stars. If white dwarfs, instead of regular
stars, are tidally disrupted by black holes
\citep{Fro94,Fry99,Ros09}, much more strongly magnetized disk with
higher viscous parameter can be produced, because during the
contraction process which forms the white dwarf, magnetic fields are
significantly amplified due to magnetic flux conservation (Tout et
al. 2004). As a matter of fact, white dwarfs can only be tidally
disrupted by IMBHs, because they will simply fall into supermassive
black holes without being tidally disrupted, due to their much
smaller sizes than regular stars. Similarly, neutron stars, though
have much higher surface magnetic fields than those of white dwarfs,
will fall into IMBHs directly without being tidally disrupted. We
thus postulate that the linear relations between $T_k$ and $T_{90}$
given in Equations (13) and (14) divide 25 GRB\,060614-type bursts
into two subclasses: 4 of them are Type I, corresponding to IMBHs
gulping to solar-type stars, and the rest 21 bursts are Type II,
which are produced by the IMBHs disrupting white dwarfs,
respectively. The ratio of Type I to Type II is about 1:5,
suggesting that white dwarfs are much more abundant than solar-type
stars around IMBHs; this possibility is discussed in the end.

\section{Discussion and conclusion}

We have identified a new class of GRBs, called GRB\,060614-type
bursts, which is produced by IMBHs tidally disrupting stars. To
select out these GRB\,060614-types, we have used criteria based on
the observation features of GRB\,060614 and our tidal disruption
model, and searched through all the GRBs discovered by {\it Swift}
BAT until 2008 October 1.

(1) We found 25 GRB\,060614-type bursts from the 328 {\it Swift}
GRBs with accurately measured durations and without SN association;
4 bursts (GRB060210, GRB060614, GRB080602, and GRB080503) are Type
I, and the rest 21 bursts are Type II.  These new GRB\,060614-type
bursts make $6\%$ of the total $\textit{Swift}$ GRBs, composed of
$1\%$ Type I and $5\%$ Type II.

(2) We derive the distributions of the IMBH's masses of the 25
bursts, as well as the isotropic energies and redshifts for the 18
bursts with unknown redshifts, predicted by the tidal disruption
model. The statistical studies show that the masses of the IMBHs are
from $5\times 10^3M_\odot$ to $9\times 10^4M_\odot$ (in the left
panel of Figure 5), the isotropic energies predicted are well
consistent with those given by \citet{Koc08}, and the redshifts
predicted are also agreed well with observations. These results
confirm the tidal disruption model, and may possibly provide a new
standard candle to study the cosmology.

(3) We obtain statistically that the relation between the
substructure period and the duration of the 25 GRB\,060614-type
bursts is fitted by two linear curves with slopes (viscous
parameters of their disks) of $0.08$ and $0.27$, respectively. This
indicates that 25 GRB\,060614-type bursts are composed of 2 subtypes
called Type I and Type II, corresponding to two different
progenitors for their productions through the tidal disruption. We
postulate that the progenitors for 4 Type I bursts are most likely
solar-type stars, and those for the 21 Type II bursts are probably
white dwarfs.

Finally, we discuss the event rate ratio between Type I and Type II
bursts. Gebhardt et al.(2005) found evidence for an IMBH of mass of
about $2\times 10^4M_\odot$, residing in a globular cluster G1,
which makes globular cluster as the most popular candidate to probe
IMBHs. In young star clusters, high-mass stars segregate through
energy equipartition; as a result, the heavier stars sink to the
center while the lighter stars move to the outer halo. This process
is called ``mass segregation'' (Spitzer 1969, 1987). Through
dynamical friction, most massive stars tend to concentrate toward
the center and drive the system to core collapse (G\"{u}rkan et al.
2004). As shown in a number of numerical studies (e.g., Portegies
Zwart et al. 2004; G\"{u}rkan et al. 2004), very high initial
central densities might lead to a rapid core collapse and
segregation of massive stars, and trigger a runaway merger of
massive stars, leading to the formation of an IMBH. It is naturally
suggested that the less massive stars, not involved in the runaway
process but were also migrating toward the clusters' centers, may
eventually end as white dwarfs, neutron stars or black holes.
Neutron stars and black holes, driven by dynamical frictions, will
also migrate toward the center , due to their larger masses, so only
white dwarfs (and some low mass stars) will remain (Heyl 2008; Heyl
\& Penrice 2009). On the other hand, due to the mass segregation
mentioned above, those much less massive stars, such as the
solar-type stars, may migrate there much later on the average.
Therefore in the vicinity of IMBHs in centers of star clusters, the
space density of white dwarfs should be much higher than that of
solar-type stars, explaining naturally the much higher event rate of
Type II bursts discussed above. We thus predict that the central
region of the G1 cluster harboring an IMBH should contain many more
white dwarfs than solar-type stars. However, without further
extensive studies, which are beyond the scope of this present work,
we cannot predict or explain quantitatively the suggested 5:1 ratio
between white dwarfs and solar-type stars.

\begin{acknowledgments}
We thank J.S. Deng for the improvement on writings. This research
was supported by the National Natural Science Foundation of China
(grants 10573021,  10821061, 10733010 and 10725313), the Chinese
Academy of Science through project no. KJCX2-YW-T03, and the 973
Program of China under grant 2009CB824800.
\end{acknowledgments}

{}

\clearpage

\begin{deluxetable}{lccccccccccc}

\tablecolumns{8} \tablewidth{0pc} \tablecaption{$Swift$ Gamma-Ray
Bursts of~GRB\,060614-type with known redshifts}

\tablehead{ \colhead{GRB} & \colhead{$T_{\rm 90}$\tablenotemark{a}}
& \colhead{$T_{\rm k}$\tablenotemark{d}}  & \colhead{$T_{\rm
pulse}$\tablenotemark{b}} & \colhead{{$(z^{obs}$)}\tablenotemark{a}}
& \colhead{$N_{\rm tot}$\tablenotemark{b}} & \colhead{$\log
(E^{KB08}_{\rm iso})$\tablenotemark{c}} &
\colhead{$M_5$\tablenotemark{b}} &
\colhead{$\alpha$\tablenotemark{b}} &  \colhead{$\log (E^{pre}_{\rm
iso})$\tablenotemark{b}} &  \colhead{Case\tablenotemark{e}}
\\
\colhead{} & \colhead{(s)} & \colhead{(s)} & \colhead{s}  &
\colhead{} & \colhead{} & \colhead{(ergs)} & \colhead{} & \colhead{}
& \colhead{(ergs)}& \colhead{} }

\startdata

050505  &  58.9  &  25  & 1.21 &  4.27   &  62   & 53.18  &  0.50  & 0.42   &53.11  &  II \\
051109a &  37.2  &  15  & 0.98 &  2.346  &  46   & 52.36  &  0.30  & 0.40   &52.78  &  II \\
060116  &  105.9 &  22  & 0.7  &  4      &  126  & 53.32  &  0.44  & 0.21   &53.67  &  II \\
060210  &  255   &  20  & 0.8  &  3.91   &  100  & 53.62  &  0.40  & 0.08   &53.95  &  II \\
060614  &  102   &   9  & 1.2  &  0.125  &   30  & 51.03  &  0.18  & 0.09   &51.75  &  II \\
060926  &  8     &  2.5 & 0.28 &  3.208  &   36  & 51.97  &  0.05  & 0.31   &52.00  &  II \\
061007  &  75.3  &  26  & 0.65 &  1.261  &  120  & 54.18  &  0.52  & 0.35   &53.50  & III \\

\enddata

\tablenotetext{a} {From $http://swift.gsfc.nasa.gov$}

\tablenotetext{b} {Estimated in this work}

\tablenotetext{c} {From \cite{Koc08}}

\tablenotetext{d} {Estimated by visually inspecting the prompt
$\gamma$-ray light curves of the samples}

\tablenotetext{e} {Classified cases of light curves in the process
of identifying sub-bursts and their durations}

\end{deluxetable}

\begin{deluxetable}{lccccccccccc}

\tablecolumns{10} \tablewidth{0pc} \tablecaption{18 Gamma-Ray Bursts
of GRB\,060614-type with unknown redshifts}

\tablehead{ \colhead{GRB} & \colhead{$T_{\rm 90}$\tablenotemark{a}}
& \colhead{$T_{\rm k}$\tablenotemark{d}} & \colhead{$N_{\rm
tot}$\tablenotemark{b}} & \colhead{$T_{\rm pulse}$\tablenotemark{b}}
& \colhead{$M_5$\tablenotemark{b}} &
\colhead{$\alpha$\tablenotemark{b}} &
\colhead{{$(z^{pre})$}\tablenotemark{b}} & \colhead{$\log
(E^{pre}_{\rm iso})$\tablenotemark{b}}&
\colhead{Case\tablenotemark{c}}
\\
\colhead{} & \colhead{(s)} & \colhead{(s)}  & \colhead{} &
\colhead{(s)} & \colhead{} & \colhead{} & \colhead{} &
\colhead{(ergs)}& \colhead{}}

\startdata

050117  & 166.6  &  45 & 141  & 0.96 &0.90  &  0.27  &3.27 &53.92  &  II \\
050306  & 158.3  &  33 & 186  & 0.89 &0.66  &  0.21  &3.31 &54.01  &  II \\
050326  & 29.3   &  13 &  19  & 2.05 &0.26  &  0.44  &0.71 &52.29  &   I \\
050607  & 26.4   &  10 &  26  & 1.15 &0.20  &  0.38  &2.32 &52.38  &  II \\
050717  & 85     &  16 &  55  & 1.16 &0.32  &  0.19  &1.93 &53.21  &  II \\
060102  & 19     &  6  &  19  & 0.95 &0.12  &  0.32  &2.61 &52.10  &  II \\
060105  & 54.4   &  20 &  57  & 1.05 &0.40  &  0.37  &1.08 &53.04  & III \\
060306  & 61.2   &  20 &  63  & 0.95 &0.40  &  0.33  &2.78 &53.13  &   I \\
060424  & 37.5   &  11 &  60  & 0.73 &0.22  &  0.29  &3.59 &52.90  &  II \\
060510a & 20.4   &  8  &  25  & 0.64 &0.16  &  0.39  &0.72 &52.25  &  II \\
061028  & 106.2  &  40 & 117  & 1.03 &0.80  &  0.38  &6.36 &53.64  &  II \\
070223  & 88.5   &  26 &  95  & 0.82 &0.52  &  0.29  &4.23 &53.47  &  II \\
080212  & 123    &  31 & 198  & 0.47 &0.62  &  0.25  &5.3 &53.93   &  II \\
080328  & 90.6   &  24 &  83  & 0.87 &0.48  &  0.26  &1.99 &53.42  & III \\
080409  & 20.2   &  8  &  21  & 0.76 &0.16  &  0.40  &1.88 &52.17  &  II \\
080503  & 170    &  15 &  33  & 0.91 &0.30  &  0.09  &3.34 &53.29  &  II \\
080602  & 74     &  5  &  33  & 0.61 &0.10  &  0.07  &1.95 &52.93  & III \\
080723a & 17.3   &  8  &  18  & 1.33 &0.16  &  0.46  &2.13 &52.04  &  II \\
\enddata

\tablenotetext{a} {From $http://swift.gsfc.nasa.gov$}

\tablenotetext{b} {Estimated in this work}

\tablenotetext{c} {Classified cases of light curves in the process
of identifying sub-bursts and their durations}

\tablenotetext{d} {Estimated by visually inspecting the prompt
$\gamma$-ray light curves of the samples}

\end{deluxetable}

\begin{figure}
\centering
\includegraphics[width=6in]{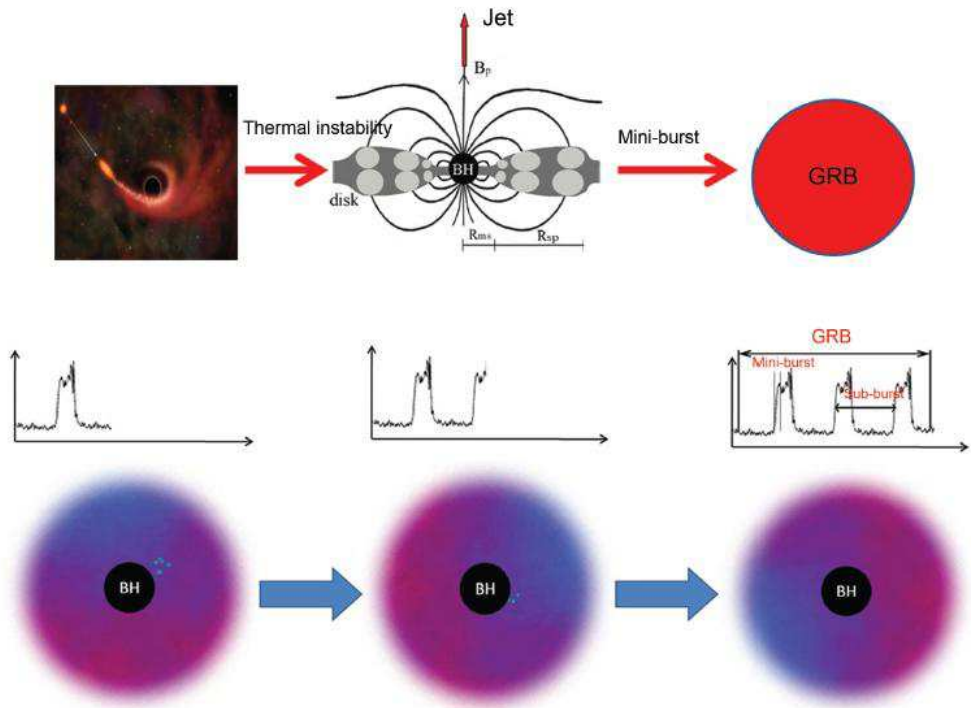}
   \caption{Scheme describing the general picture of the
tidal disruption model for GRB 060614. The upper panel shows an IMBH
gulping a solar-type star and triggers an intense blast of gamma
rays. The lower panel emulates the above processes using a flash.}
       \end{figure}

\begin{figure}
\centering
\includegraphics[width=6in]{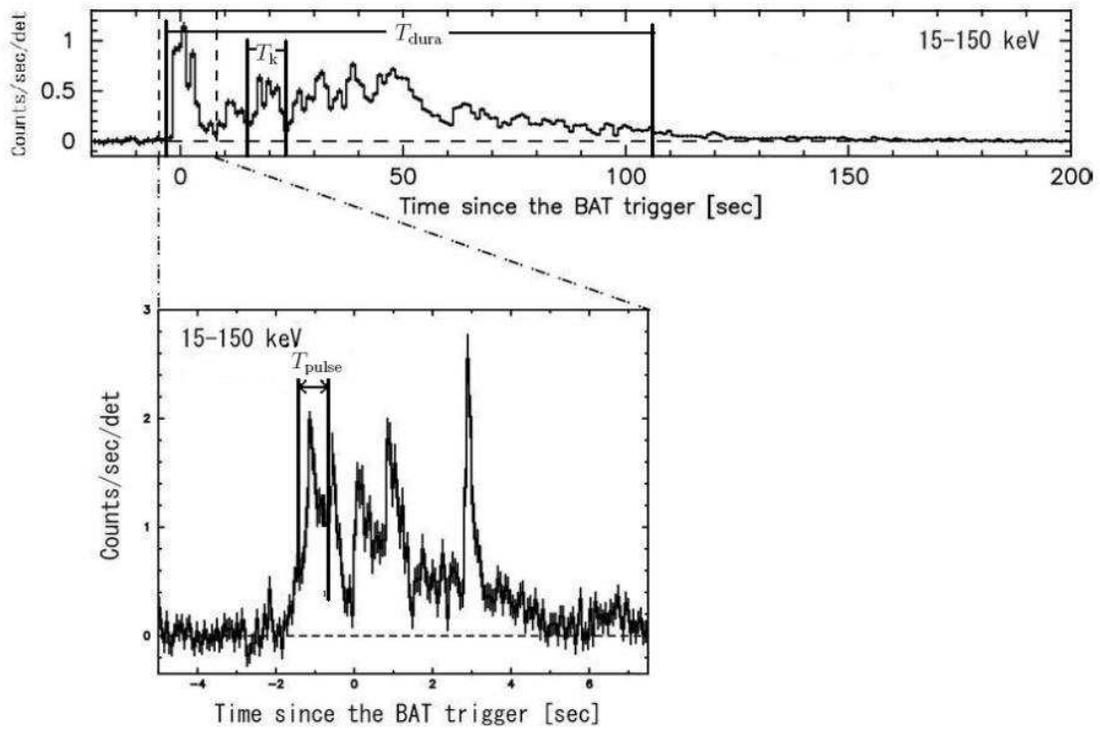}
   \caption{Observed timescales in the light curve of GRB 060614 in the tidal disruption model: $T_{\rm dura}$ corresponds to the duration,
    $T_k$ is the periodicity of the sub-burst, and $T_{\rm pulse}$ corresponds to the mini-burst timescale.}
       \end{figure}

\bigskip

\begin{figure}
 \centering
\includegraphics[width=6in]{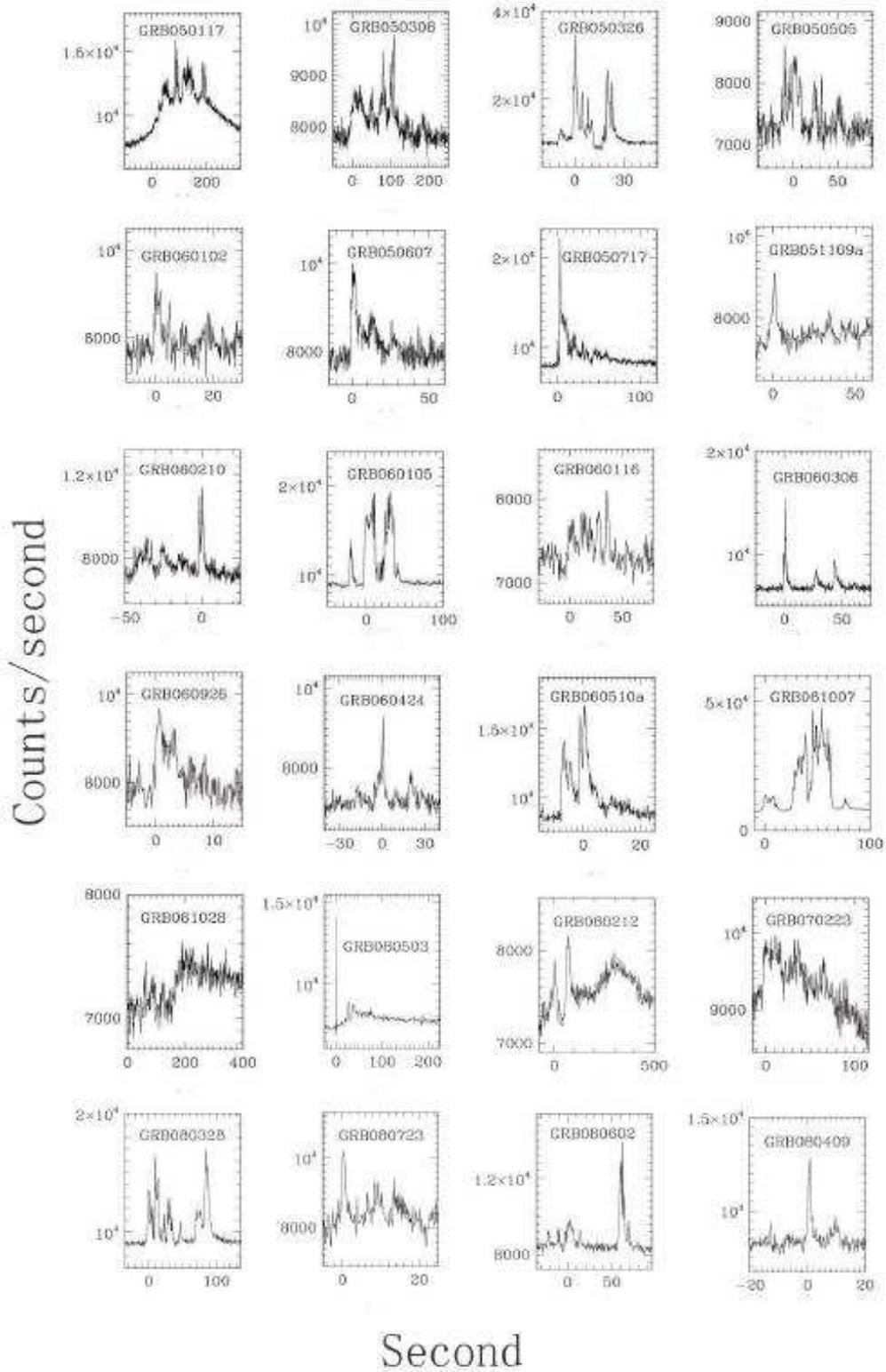}
\caption{BAT light curves of all 24 selected GRB\,060614-types
downloaded from
$http://swift.gsfc.nasa.gov/docs/swift/archive/grb\_table.html$.}
\end{figure}

\begin{figure}
 \centering
\includegraphics[width=6in]{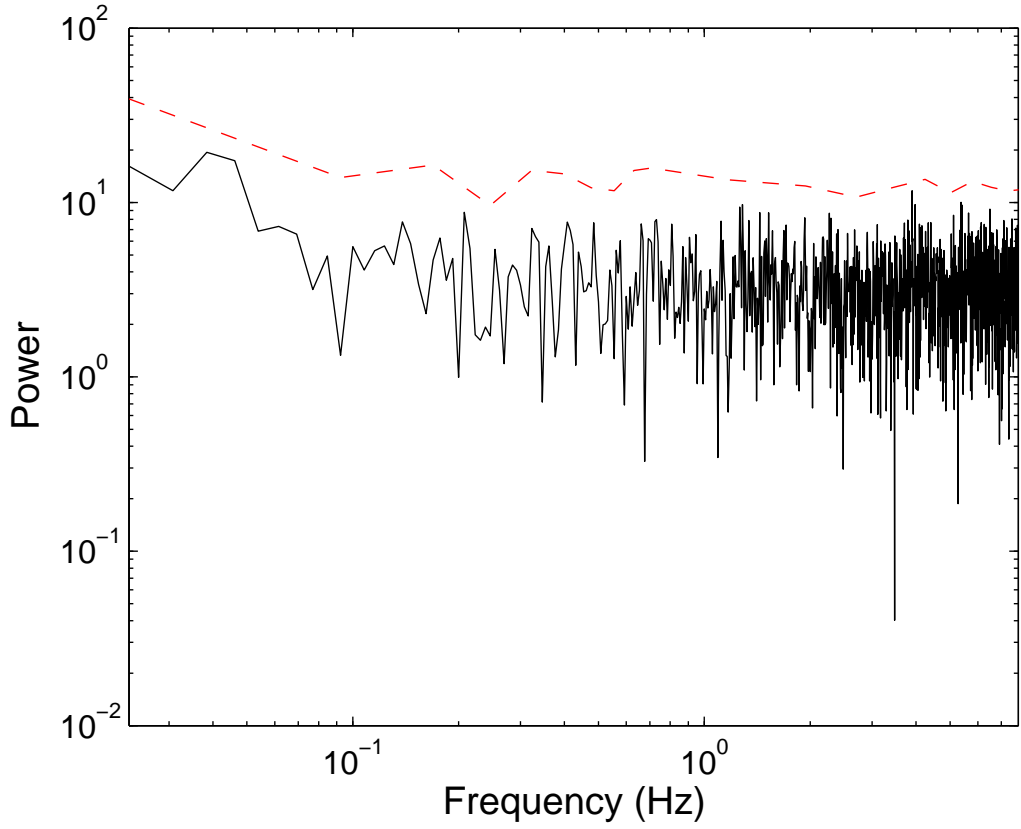}
   \caption{Power spectrum of the 64 ms BAT light curve in the
15-350 keV band from $T_{\rm -20}$\,s to $T_{\rm +110}$\,s of
GRB\,070223. The red line denotes the threshold for the detection of
sinusoidal signals at the $3\sigma$ confidence level.}
      \end{figure}

\begin{figure}
\centering
\includegraphics[width=6in]{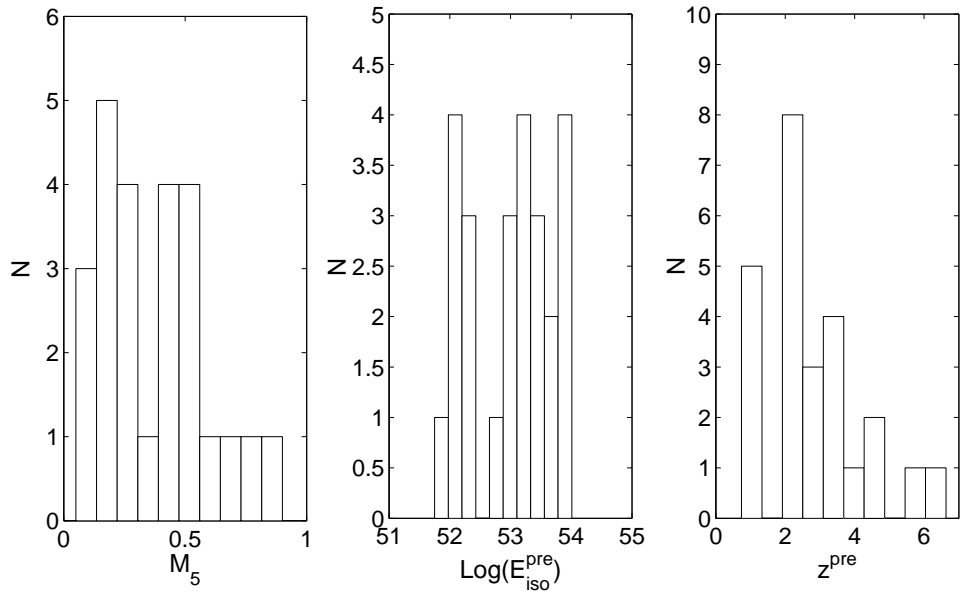}
  \caption{Distributions for the black hole
masses, the isotropic energy, and the redshifts for 25
GRB\,060614-type bursts. $N$ is the number of GRBs. }
\end{figure}

\begin{figure}
 \centering
\includegraphics[width=6in]{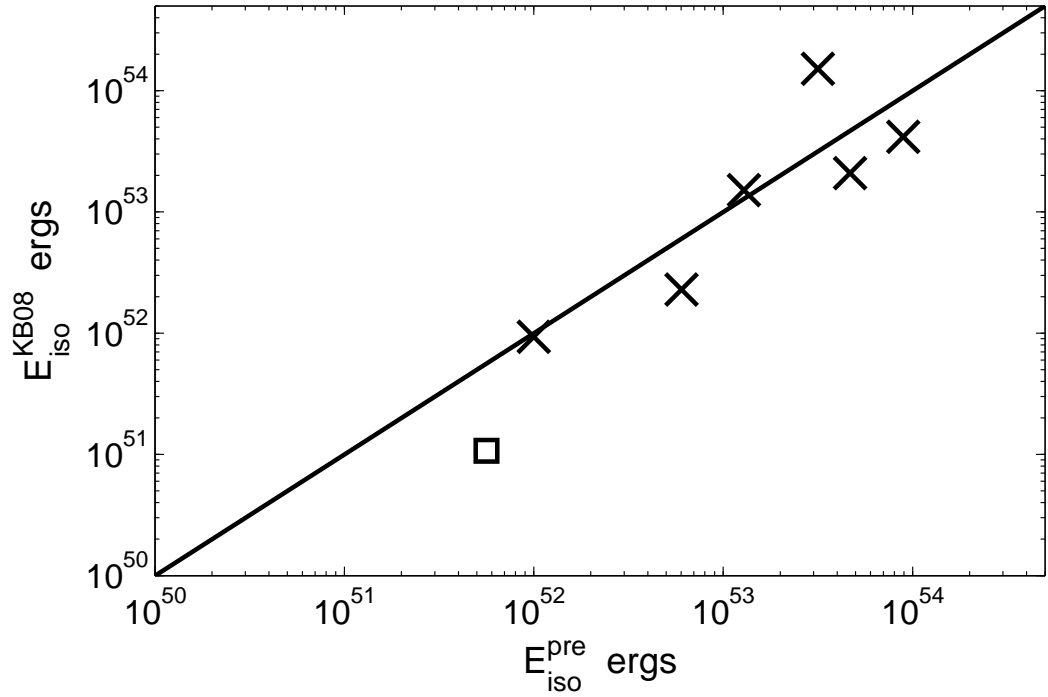}
   \caption{Isotropic energy  from \citet{Koc08} compared with
   those predicted by this model for seven known redshift samples. The square refers to
GRB\,060614, and the crosses stand for the six GRB 060614-type
bursts. The solid line refers to the least-squares fitting:
$\log(E_{\rm iso}^{\rm KB08})=(0.997\pm 0.03)\log(E_{\rm iso}^{\rm
pre})$ with the adjusted $\Re$-square of 0.999.}
      \end{figure}

\begin{figure}
\centering
\includegraphics[width=6in]{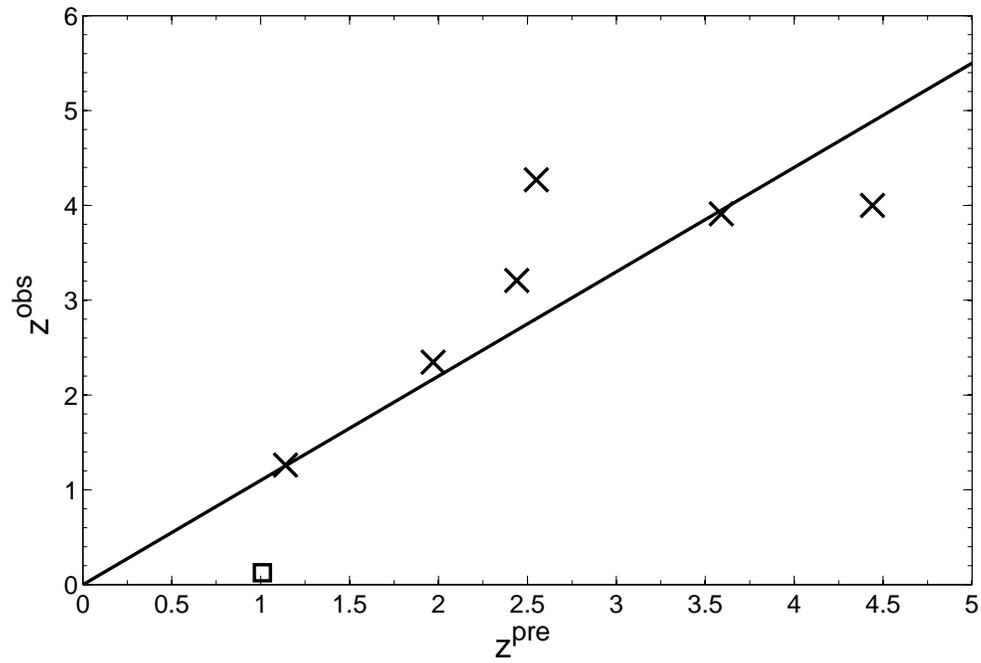}
    \caption{Redshifts from observations compared with the redshifts given by this model.
The square refers to GRB\,060614, and the crosses stand for the six
GRB 060614-type bursts. The solid line refers to the least-squares
fitting: $z^{\rm obs}=(1.10\pm 0.12)z^{\rm pre}$ with the adjusted $
\Re$-square of 0.93.}
\end{figure}

\begin{figure}
\centering
\includegraphics[width=6in]{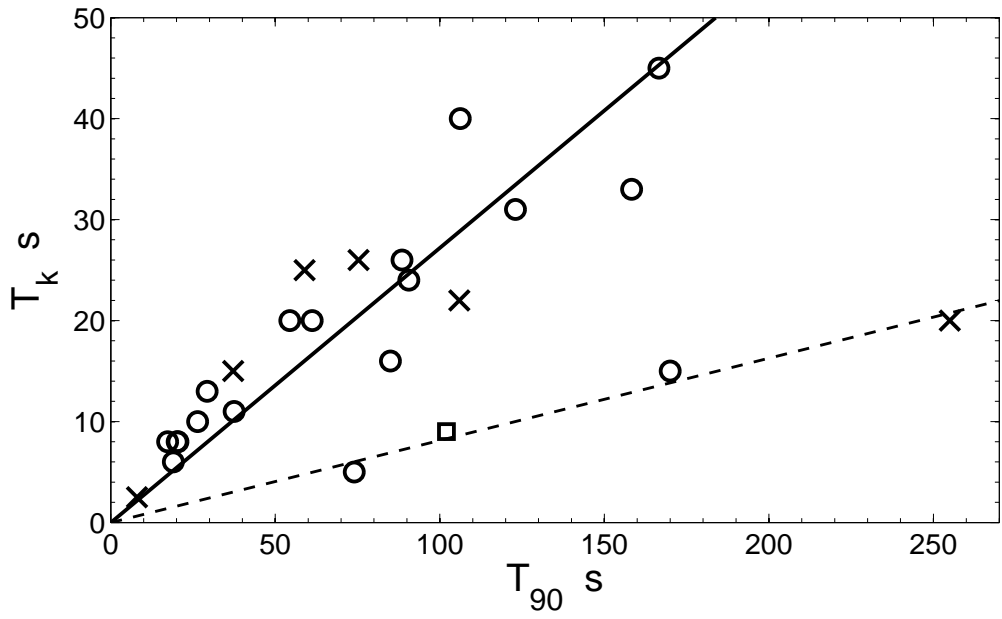}
   \caption{Relation between the GRB duration
and the substructure period: the square refers to GRB\,060614 , the
crosses denote the known redshift samples, and the circles denote
the unknown redshift samples. The dashed and solid lines are the
least-squares fitting for two different slopes, which correspond to
two different viscous parameters of the disk.}
\end{figure}


%
%
%
%
%
%
%
%

\end{document}